# Natural sinks and sources of $CO_2$ and $CH_4$ in the atmosphere of Russian regions and their contribution to climate change in the 21st century as calculated with CMIP6 model ensemble

Denisov S.N.[1], Mokhov I.I.[1,2]
[1]A.M. Obukhov Institute of Atmospheric Physics RAS
[2]Lomonosov Moscow State University

denisov@ifaran.ru

**Abstract**

The natural fluxes of $CO_2$ and $CH_4$ into the atmosphere from the territory of Russia in the 21st century have been analyzed using the results of calculations with the ensemble of global climate models of the international project CMIP6. Estimates of natural $CO_2$ fluxes for Russian regions differ greatly for different models. Their values for the beginning of the 21st century range from -1 to 1 GtC/yr. In the 21st century the differences in model estimates of fluxes grow and at the end of the 21st century under the scenario with the largest anthropogenic impacts SSP5-8.5 are in the range from -2.5 to 2.5 GtC/year. Estimates of natural methane emissions to the atmosphere from the territory of Russia also differ greatly for different models. Present-day methane emissions are estimated in the range from 10 to 35 $MtCH_4$/year, while the growth in the 21st century may reach 300%. Ensemble model calculations show general trends for changes in natural greenhouse gas fluxes. Most CMIP6 ensemble models are characterized by a maximum of $CO_2$ uptake by terrestrial ecosystems and its further reduction by the end of the 21st century, while natural methane emissions to the atmosphere for all models and scenarios of anthropogenic forcing grow throughout the 21st century. The cumulative temperature potential of natural $CO_2$ fluxes on the territory of Russia in the 21st century is estimated, depending on the scenario of anthropogenic impacts, from -0.3 to 0.1 K, and the warming-accelerating impact of natural $CH_4$ emissions is estimated in the range of 0.03-0.09 K.

**Introduction**

The concentration of carbon dioxide in the Earth's atmosphere has increased one and a half times since the beginning of the industrial era from a level of about 280 ppm (Joos and Spahni, 2008) to a level of about 420 ppm by 2023 (https://gml.noaa.gov/ccgg/trends/global.html). Initially, the anthropogenic increase in atmospheric $CO_2$ concentrations was mainly due to carbon emissions from deforestation and other land-use activities. And since the mid-20th century, anthropogenic emissions from fossil fuel have become the dominant factor, and their relative share has continued to increase up to the present. Anthropogenic emissions occur against the background of natural carbon exchange, in which carbon circulates between the reservoirs of the atmosphere, ocean, biosphere and soil on scales ranging from intra-day to millennial and more (Archer et al., 2009).

According to forecast estimates, atmospheric $CO_2$ concentrations by 2100 may reach levels between 795 and 1145 ppm under the RCP 8.5 scenario of anthropogenic $CO_2$ emissions. The range of uncertainty is due to variability in terrestrial carbon exchange and uncertainty in the carbon cycle-climate feedback (Friedlingstein et al., 2014). A possibly more accurate understanding of terrestrial ecosystem response to climate change is needed to refine such quantitative estimates. This is particularly relevant in view of the Paris Agreement (2015) of the United Nations Framework Convention on Climate Change, dealing with the challenges of reducing greenhouse gas emissions and related adaptation.

The carbon balance in Russian regions with a large extent of forests, peatlands and wetlands and significant soil carbon stocks plays an important role in the global carbon cycle. Most of Russia's territory is located in the permafrost zone. In these regions, microbial decomposition of carbon is suppressed at low temperatures, while its flux from the atmosphere through photosynthesis can remain high during spring and summer seasons. Therefore, over the past millennia, large carbon stocks have accumulated in lake sediments and in bog, forest and tundra soils in these regions, which may destabilize with further warming.

Current estimates of the increase in total terrestrial $CO_2$ uptake are mainly associated with the Northern Hemisphere (Ciais et al., 2019), with boreal and temperate forests likely contributing the most (Tagesson et al., 2020). Increased temperature and lengthening of the growing season as a result of climate change should contribute to increased biospheric activity at high latitudes (Lucht et al., 2002; Chen et al., 2006). At the same time, increased "soil respiration" and growth of methane emissions into the atmosphere may compensate for the growth of carbon sink from the atmosphere associated with photosynthesis (Piao et al., 2008; Parmentier et al., 2011).

In (Denisov et al., 2019), estimates of changes in the natural fluxes of $CO_2$ and $CH_4$ to the atmosphere from the territory of Russia in the 21st century and their possible contribution to climate change were obtained using the Earth System Model of the Obukhov Institute of Atmospheric Physics of the Russian Academy of Sciences (IAP RAS ESM). It was shown that $CO_2$ uptake by terrestrial ecosystems of Russia under all considered scenarios of anthropogenic forcing reaches a maximum by the middle of the 21st century and then decreases. At the same time, natural methane emissions into the atmosphere are increasing throughout the 21st century and their contribution to the increase in global atmospheric temperature may exceed the effect of $CO_2$ uptake by natural ecosystems, so that the total effect of natural fluxes of these gases may accelerate warming by the end of the 21st century. At the same time, both methane (Melton et al., 2013; Saunois et al., 2020) and carbon dioxide (Friedlingstein et al., 2022) fluxes have high variability and their estimates from calculations with different models differ greatly even for the modern period.

This paper presents quantitative estimates of anthropogenic and natural fluxes of carbon dioxide and methane for terrestrial ecosystems of Russia in the 21st century under different scenarios of anthropogenic forcing and their contribution to global warming using the results of calculations with an ensemble of earth system models.

**Methods**

In the presented work, the analysis of natural $CO_2$ and $CH_4$ fluxes to the atmosphere from the territory of Russia in the 21st century was carried out using the results of calculations with the ensemble of global climate models of the international project CMIP6 (Eyring et al., 2016). In the previous generation of models within the CMIP5 project (Taylor et al., 2012), the results of calculations of $CO_2$ fluxes were available only for a few models, and the results of calculations of natural methane emissions were not available. CMIP6 family of models presents model estimates of $CO_2$ fluxes for dozens of earth system models and for several models estimates of $CH_4$ fluxes are available.

The analysis considered the results of model calculations under 4 scenarios of anthropogenic impacts of the SSP family (Riahi et al., 2017) in the 21st century: SSP1-2.6, SSP2-4.5, SSP3-7.0, and SSP5-8.5. Only the results of model calculations available for the period 1990-2100 were analyzed (for 1990-2015, the fluxes calculated under the "historical" scenario were considered). Information about the models, the calculations with which were used in the analysis, is given in Table 1a,b. It should be noted that some models are presented in several versions. The differences between the different model versions are related to different spatial resolution (EC-Earth3-Veg, NorESM2) or to the inclusion of the atmospheric chemistry block. It is further shown that these differences have little effect on the results obtained. In addition, the CMCC, NorESM and TaiESM1 models use earlier versions of the CLM terrestrial processes block developed for the CESM model to calculate greenhouse gas fluxes.

Table 1a. Information on model calculations used in the analysis of $CO_2$ fluxes

| Model | Institute | Land scheme | Resolution (lat x lon) | SSP scenarios 1-2.6 | 2-4.5 | 3-7.0 | 5-8.5 |
|---|---|---|---|---|---|---|---|
| ACCESS-ESM1-5 | CSIRO | CABLE 2.4 | 145x192 | 40 | 40 | 40 | 40 |
| BCC-CSM2-MR | BCC | BCC-AVIM2 | 160x320 | 1 | 1 | 1 | 1 |
| CanESM5 | CCCma | CLASS 3.6 /CTEM 1.2 | 64x128 | 50 | 50 | 50 | 50 |
| CESM2 | NCAR | CLM 5 | 192x288 | 3 | 3 | 3 | 3 |
| CESM2-WACCM | | | 192x288 | 1 | 3 | 1 | 3 |
| CMCC-ESM2 | CMCC | CLM 4.5 | 192x288 | 1 | 1 | 1 | 1 |
| CMCC-CM2-SR5 | | | 192x288 | 1 | 1 | 1 | 1 |
| CNRM-ESM2-1 | CNRM | Surfex 8.0c | 128x256 | 5 | 10 | 5 | 5 |
| EC-Earth3-Veg | Ec-Earth-Consortium | LPJ-GUESS | 256x512 | 7 | 8 | 5 | 6 |
| EC-Earth3-Veg-LR | | | 160x320 | 3 | 3 | 3 | 3 |
| EC-Earth3-CC | | | 256x512 | - | 1 | - | 1 |
| GFDL-ESM4 | NOAA-GFDL | GFDL-LM4.1 | 180x288 | 1 | - | - | - |
| IPSL-CM5A-INCA | IPSL | ORCHIDEE | 96x96 | 1 | - | 1 | - |
| IPSL-CM6A-LR | | | 143x144 | 6 | 11 | 11 | 6 |
| MPI-ESM1-2-LR | MPI-M | JSBACH 3.20 | 96x192 | 10 | 10 | 10 | 10 |
| MRI-ESM2-0 | MRI | HAL 1.0 | 160x320 | - | - | - | 1 |
| NorESM2-LM | NCC | CLM | 96x144 | 1 | 3 | 1 | 1 |
| NorESM2-MM | | | 192x288 | 1 | 2 | 1 | 1 |
| TaiESM1 | AS-RCEC | CLM 4.0 | 192x288 | 1 | 1 | 1 | 1 |

Table 1b. Information on model calculations used in the analysis of $CH_4$ fluxes

| Model | Institute | Land scheme | Resolution (lat x lon) | SSP scenarios 1-2.6 | 2-4.5 | 3-7.0 | 5-8.5 |
|---|---|---|---|---|---|---|---|
| CESM2 | NCAR | CLM 5 | 192x288 | 3 | 3 | 3 | 3 |
| CESM2-WACCM | | | 192x288 | 1 | 3 | 1 | 3 |
| NorESM2-LM | NCC | CLM | 96x144 | 1 | 3 | 1 | 1 |
| NorESM2-MM | | | 192x288 | 1 | 2 | 1 | 1 |
| UKESM1-0-LL | MOHC | JULES-ES-1.0 | 144x192 | 16 | 5 | 16 | 5 |

Tables 1a,b summarize the number of numerical model calculations under different initial conditions for each SSP scenario. In this paper, the average values of fluxes for all model variants of calculations for each scenario were used in the analysis.

In this paper we analyzed the data available on the website (https://esgf-node.llnl.gov/projects/cmip6/), including the variables "NEP" (net ecosystem production) and "wetlandCH4" to determine the natural fluxes of $CO_2$ and $CH_4$. However, NEP data for CESM2, CESM2-WACCM and IPSL-CM5A-INCA models seem to be presented with the opposite sign (emissions rather than $CO_2$ uptake), so these data were used in the analysis in this paper with the opposite sign.

For comparison, calculations were performed with the IAP RAS EMS (Mokhov and Eliseev 2012; Eliseev et al. 2014; Denisov et al. 2015) (see also (Mokhov et al. 2002; Mokhov et al. 2005; Eliseev and Mokhov 2011)). The IAP RAS EMS belongs to the class of global climate models of intermediate complexity (Claussen et al., 2002; Eby et al., 2013; Zickfeld et al., 2013; MacDougall et al., 2020). Large-scale dynamics of the atmosphere and ocean are described explicitly in the model, and synoptic processes are parameterized, which allows us to significantly increase the speed of calculations. The model contains a block of the carbon cycle, including the methane cycle, which takes into account emissions into the atmosphere and absorption of carbon dioxide and methane by different natural ecosystems (Eliseev et al. 2008; Denisov et al. 2013).

Numerical calculations were carried out with the IAP RAS ESM with 40 and 60 model cells in latitude and longitude resolution and an integration step of 5 days. Numerical calculations for the period 1850-2100 under different scenarios of anthropogenic forcing were carried out. Changes in the content of greenhouse gases, tropospheric and stratospheric volcanic sulfate aerosols in the atmosphere, changes in the total solar radiation at the upper boundary of the atmosphere and changes in the area of agricultural land were taken into account. For the period 1850-2014, these forcings were set according to the "historical" scenario of the CMIP6 project. For the period 2015-2100 anthropogenic forcing was set according to the scenarios of the SSP family.

The contribution of natural fluxes of $CO_2$ and $CH_4$ in the atmosphere in Russian regions to global climate change was estimated using the cumulative temperature potential CT similarly (Denisov et al., 2019, 2022). The cumulative climate effect of the greenhouse gas source can be estimated on the time horizon $[T_0;T_H]$ as follows:

$$CT(T_0, T_H) = \sum_{t=T_0}^{T_H-1} E(t) GTP^{(a)*}(t, T_H), \qquad (1)$$

where E(t) is the greenhouse gas flux for year t, and $GTP^{(a)*}(t,T_H)$ is the absolute global temperature change potential associated with a particular gas at the horizon $[t;T_H]$, modified to account for changes in background conditions. A detailed description of the GTP modification and its effects is given in (Denisov et al., 2022).

**Results**

Fig.1 shows the CMIP6 ensemble average of $CO_2$ uptake by terrestrial ecosystems in Northern Eurasia with inter-model standard deviations for the period 1990-2014. On average, net annual natural $CO_2$ uptake from the atmosphere exceeds atmospheric emissions over most of the Russian territory with characteristic values of 20-40 $gC/m^2/year$, which agrees with the estimates (Friedlingstein et al., 2022). $CO_2$ absorption has a pronounced zonal character and its maximums are reached in the European territory of Russia in the band 50-60°N. Practically for the entire territory of Russia, the standard deviation of these estimates significantly exceeds the $CO_2$ absorption values with characteristic values of 40-70 $gC/m^2/year$.

Fig. 2 shows the annual average total natural fluxes of $CO_2$ from terrestrial ecosystems to the atmosphere in Russia (net ecosystem production, NEP). Negative values in the graphs correspond to the absorption of $CO_2$ from the atmosphere. Due to the high variability of fluxes, the results of calculations with the IAP RAS EMS are presented as a 9-year moving average.

Estimates of total annual average natural $CO_2$ fluxes for the Russian regions vary greatly for different models. Their present-day values (for the beginning of the 21st century, up to 2014, in accordance with the "historical" scenario) range from -1 to 1 GtC/year. At the same time, for most models (excluding BCC, IPSL, CNRM and MRI models) the range of modern values of $CO_2$ fluxes is much narrower: from -0.5 to -0.2 GtC/yr. In the 21st century, the divergence in model flux estimates is growing. The largest range of estimates from -2.5 to 2.5 GtC/year is reached at the end of the 21st century under the scenario with the largest anthropogenic forcing SSP5-8.5. The previously mentioned models give a range of estimates from -1 to -0.2 GtC/year throughout the 21st century (excluding the CanESM5 model under more aggressive anthropogenic scenarios).

In (Dolman et al., 2012), estimates of atmospheric carbon uptake by terrestrial ecosystems in Russia using various methods are presented. Estimates based on dynamic models of processes in terrestrial ecosystems (similar to CMIP6 model blocks) showed average values of about 0.2 GtC/year and a high inter-model discrepancy of 100% of this value. Using other methods, estimates of $CO_2$ uptake in Russian regions are significantly higher: 0.6-0.8 GtC/yr (Gurney et al., 2003; Dolman et al., 2012; Ciais et al., 2010).

For comparison, Fig. 2 shows estimates of $CO_2$ fluxes according to calculations with the IAP RAS ESM and anthropogenic $CO_2$ emissions from the territory of Russia under the corresponding scenarios. In (Denisov et al., 2019, 2022) anthropogenic emissions from the territory of Russia were calculated according to the RCP family scenarios (Moss et al., 2010) for the REF region (countries of Eastern Europe and the former USSR) with normalization to modern values for Russia. For the SSP family scenarios, anthropogenic emission values for individual countries, including Russia, are available.

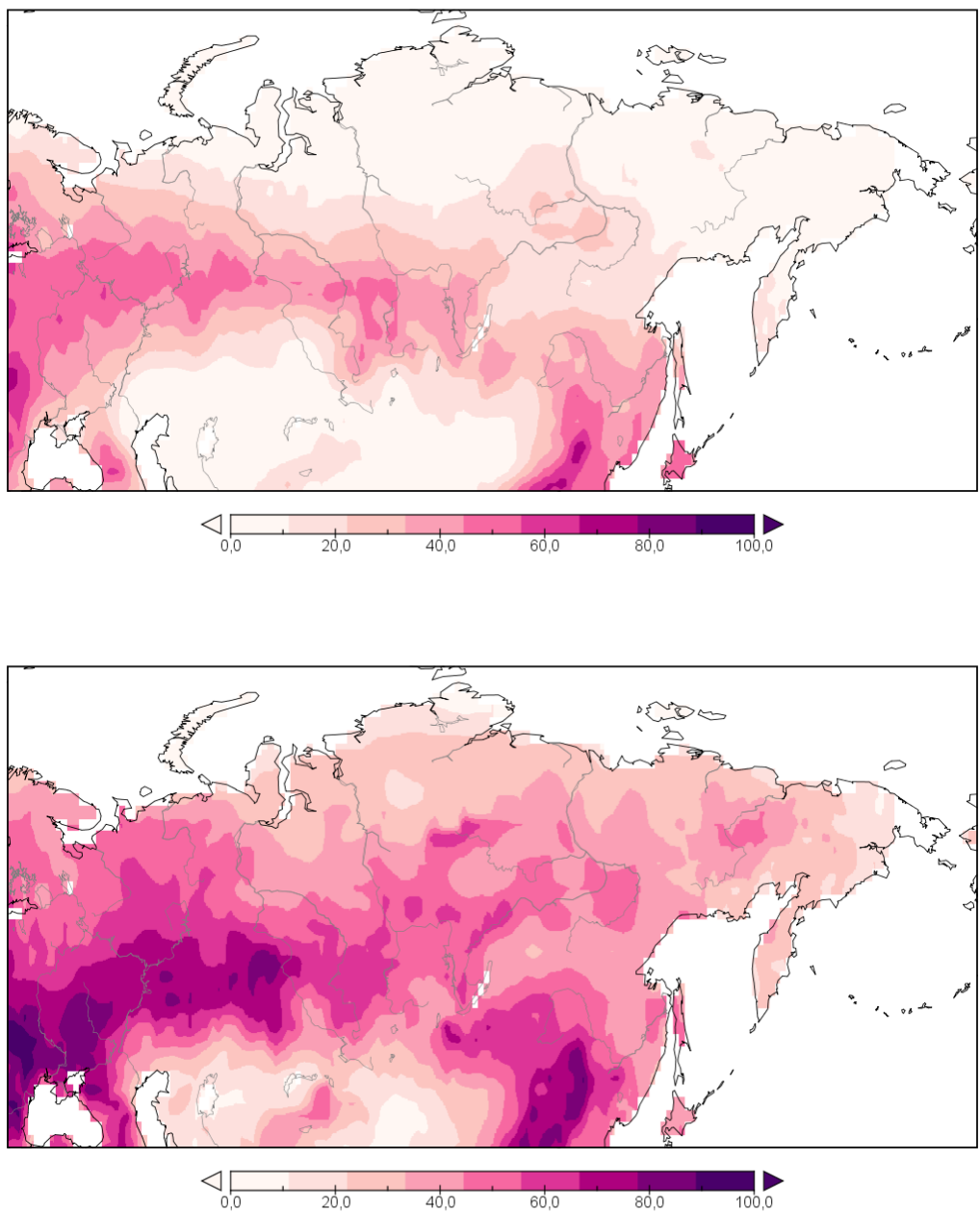

Fig. 1. Average $CO_2$ uptake (NEP) by terrestrial ecosystems of Northern Eurasia [$gC/m^2/year$] (top) and standard deviation of these estimates (bottom).

Estimates of natural $CO_2$ fluxes using IAP RAS ESM correspond to the range of estimates using CMIP6 ensemble models throughout the 21st century under all analyzed scenarios of anthropogenic impacts. If the BCC model is excluded, the IAP RAS ESM estimates are slightly out of the range of estimates of most of the considered CMIP6 ensemble models only under the

SSP1-2.6 scenario at the end of the 21st century. The tendency of carbon dioxide absorption reduction by the end of the 21st century, obtained in calculations with IAP RAS ESM, is typical for many other CMIP6 models. At the same time, the higher anthropogenic emissions of greenhouse gases into the atmosphere, the later the maximum of $CO_2$ absorption is reached and its reduction begins.

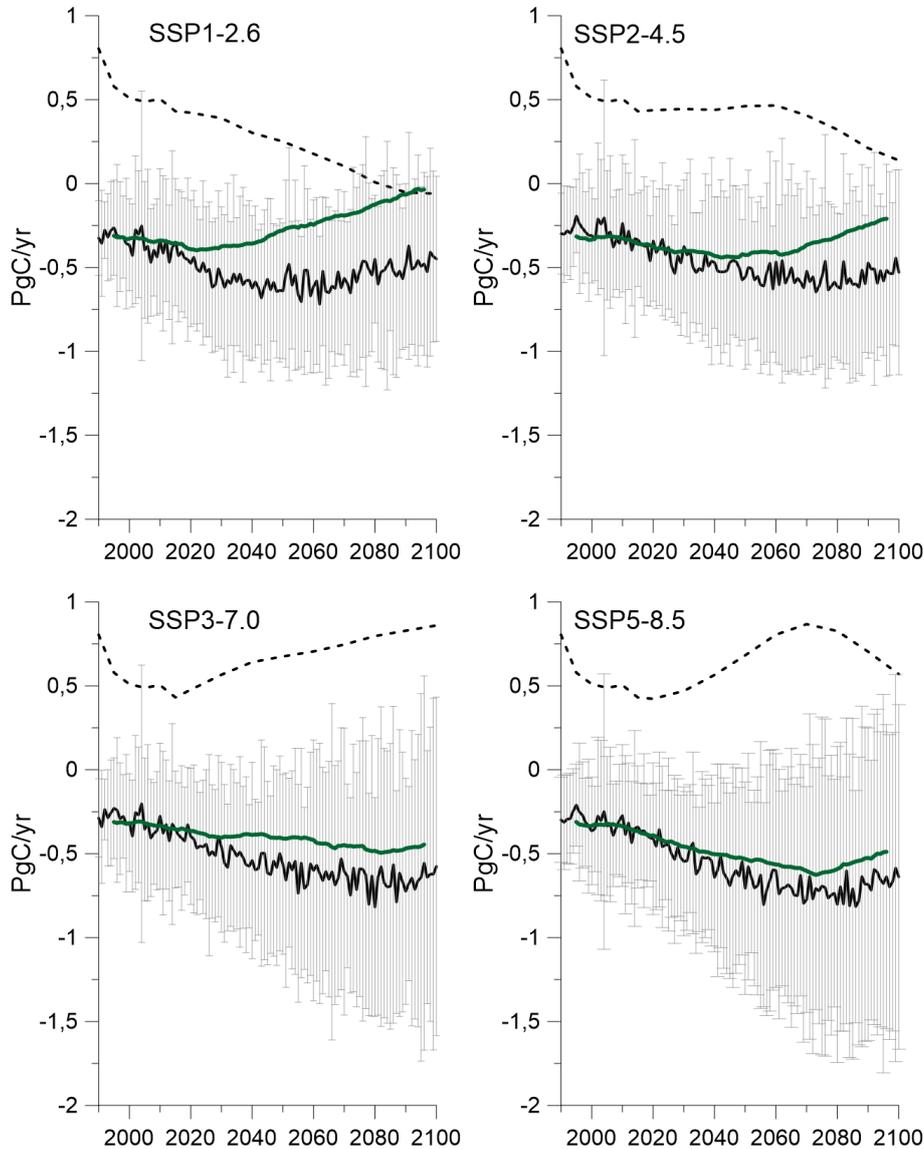

Fig. 2. Natural $CO_2$ fluxes from the atmosphere in Russian regions calculated with the ensemble of models (mean values and standard deviation) in comparison with $CO_2$ fluxes calculated with IAP RAS ESM (green line) and anthropogenic emissions into the atmosphere (dotted line) under different scenarios of anthropogenic impacts for the 21st century.

Anthropogenic $CO_2$ emissions to the atmosphere from the territory of Russia are comparable to natural fluxes from the atmosphere and can be largely compensated by them. Even under the most unfavorable scenarios, the emission values remain within the range from 0.5 to 1 GtC/year. It should be noted that the RCP 8.5 scenario corresponded to significantly higher (up to 2-2.5 GtC/year) values of anthropogenic $CO_2$ emission from the territory of Russia than the SSP5-8.5 scenario, with the same values of global radiative forcing by the end of the 21st century (8.5 W/m$^2$) (Denisov et al., 2019).

As can be seen, the BCC model differs from other models of the CMIP6 project by positive values of the natural $CO_2$ flux to the atmosphere. At the same time, modulo its difference from the ensemble mean values approximately corresponds to the models with the highest values of $CO_2$ uptake (CNRM, GFDL, IPSL).

Carbon accumulation in terrestrial ecosystems, reflected in NEP, is determined by the balance of intensities of GPP, autotrophic and heterotrophic respiration and emissions from fires:

$$NEP = GPP - R_a - R_g - F_{fire}. \qquad (2)$$

Each component in the models is determined by nonlinear dependences on a set of parameters and modulo may exceed the NEP balance value. This is associated with high interannual variability of $CO_2$ fluxes. To understand the differences between the models, it is of interest to consider the data on a finer time scale.

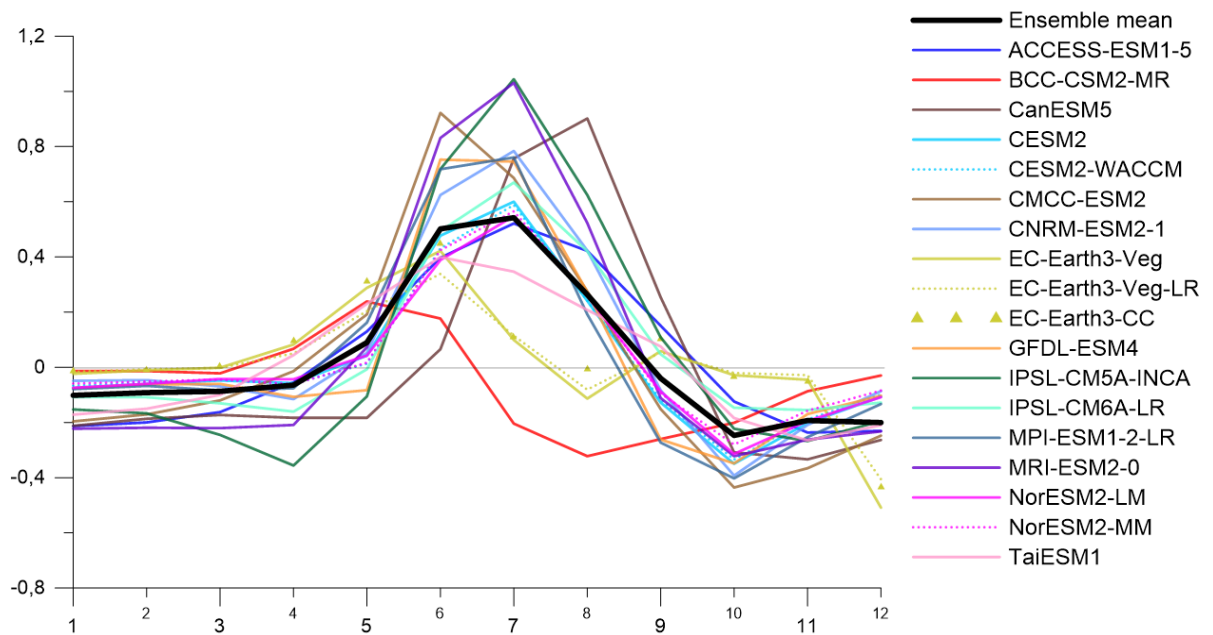

Fig. 3. Annual course of $CO_2$ uptake by terrestrial ecosystems of Russia [GtC/month] according to model calculations for the modern period (2010-2014).

Fig. 3 presents the annual course of $CO_2$ uptake by terrestrial ecosystems of Russia according to calculations with CMIP6 ensemble models for the modern period (2010-2014). On average for the ensemble, $CO_2$ uptake occurs from May to September with values up to 0.5 GtC/month, and in the remaining months there are $CO_2$ emissions from terrestrial ecosystems with intensity up to 0.3 GtC/month. For individual models, the range of total $CO_2$ fluxes is -0.5 to 1.1 GtC/month. For most models, the maximum of $CO_2$ uptake is reached in July and the maximum of $CO_2$ emission to the atmosphere in October-November. At the same time, in the BCC, CMCC, and EC-Earth3 models the absorption maximum is reached earlier, and in CanESM5 later.

The largest differences from the average for the ensemble of changes of natural CO2 fluxes in the annual variations for the Russian regions are noted for the BCC and EC-Earth3 models. The absorption maximum in them is reached in May-June, and the transition from absorption to emission to the atmosphere occurs already in July-August. (It should be noted that in some years $CO_2$ fluxes according to calculations with the BCC model can only slightly differ from other models and show the total natural absorption of $CO_2$ in the Russian regions). At the same time, although EC-Earth3 model does not stand out against the background of most other CMIP6 models in terms of mean annual values of natural $CO_2$ fluxes, it is similar to the BCC model in terms of peculiarities of the annual course of natural $CO_2$ fluxes.

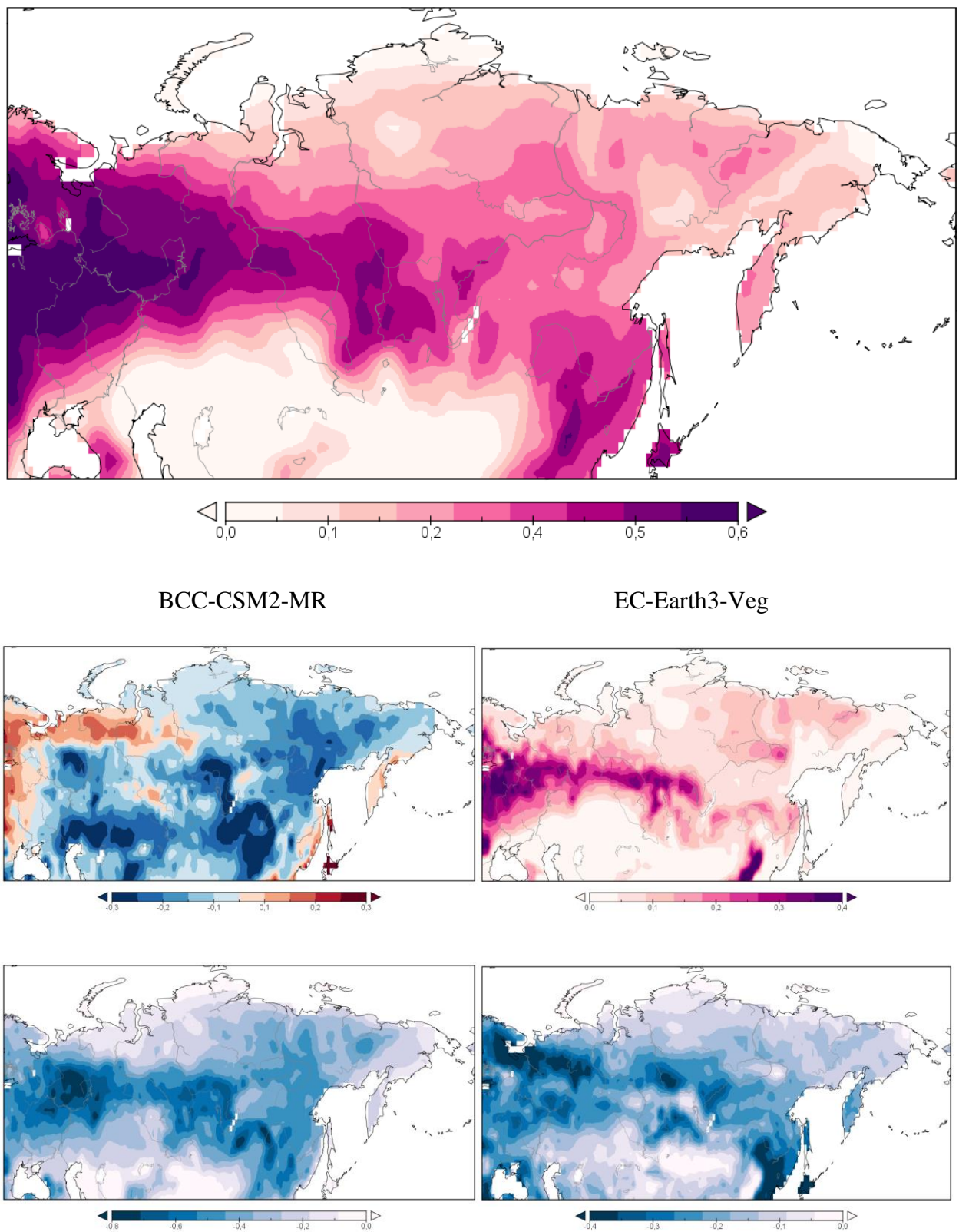

BCC-CSM2-MR EC-Earth3-Veg

Fig. 4. Mean modern $CO_2$ uptake [kgC/m$^2$/year] in Russian regions in summer by calculations with CMIP6 ensemle (top), separately by calculations with BCC and EC-Earth3 models (middle row), and their difference with the ensemble mean (bottom row).

Fig.4 presents the results of calculations of modern summer values of natural $CO_2$ fluxes on average for the CMIP6 model ensemble and separately for calculations with the BCC and EC-Earth3 models. On average for the ensemble of models, the maximum $CO_2$ uptake by natural ecosystems was observed for the western and central parts of the European territory of Russia. The spatial distribution of $CO_2$ fluxes in the EC-Earth3 model is similar to the ensemble average, but the characteristic values of fluxes over the whole territory of Russia are twice smaller. According to calculations with the BCC model, $CO_2$ is released into the atmosphere from the most part of the Russian territory in summer. Absorption is noted only in the northern and western parts of the European territory, in Primorsky Krai and Kamchatka. $CO_2$ emission to the atmosphere reaches maximum values in the southern part of Siberia, with the largest anomalies compared to the ensemble mean estimates for the ETR.

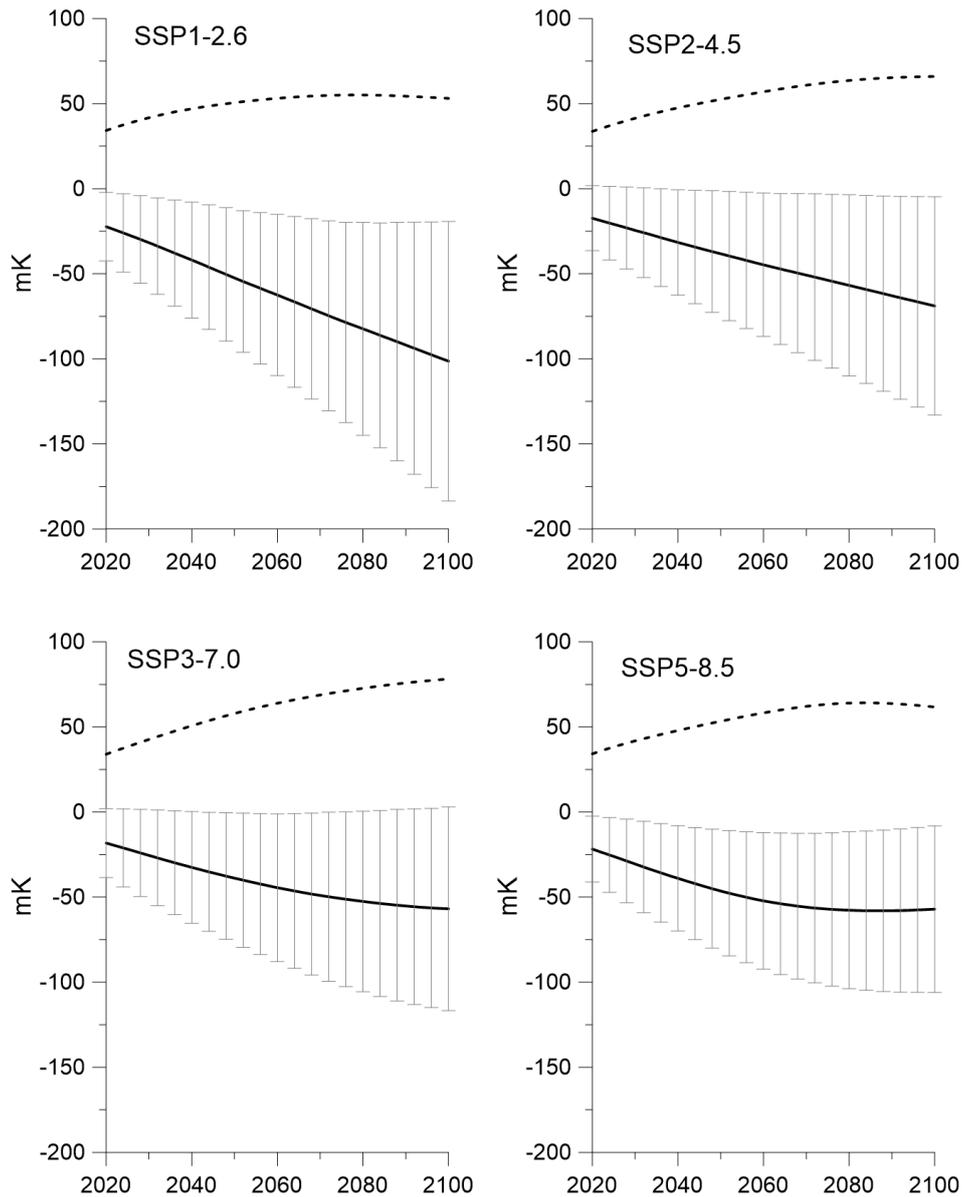

Fig. 5. Cumulative temperature potential of natural $CO_2$ fluxes [mK] on the territory of Russia since 1990 as calculated with ensemble models (mean values and standard deviation) and anthropogenic fluxes (dashed line) under different scenarios of anthropogenic forcing.

The cumulative temperature potential of natural $CO_2$ fluxes on the territory of Russia from 1990 to the end of the 21st century is estimated in the range from -0.3 to 0.1 K, depending on the scenario of anthropogenic impacts. Its average over the ensemble of models is about -0.1 K for the

SSP1-2.6 scenario and about -0.06 K for the other scenarios (Fig. 5). As in the case of $CO_2$ fluxes, the range of estimates of the cumulative temperature potential is reduced when the core group of models is isolated. The tendency noted earlier for the IAP RAS ESM to slow down the growth and even to weaken the stabilizing contribution to global climatic changes in terrestrial ecosystems of Russian regions in the 21st century (Denisov et al., 2019) is also noted for most CMIP6 models under scenarios with strong anthropogenic impacts even without taking into account the contribution of natural methane emissions.

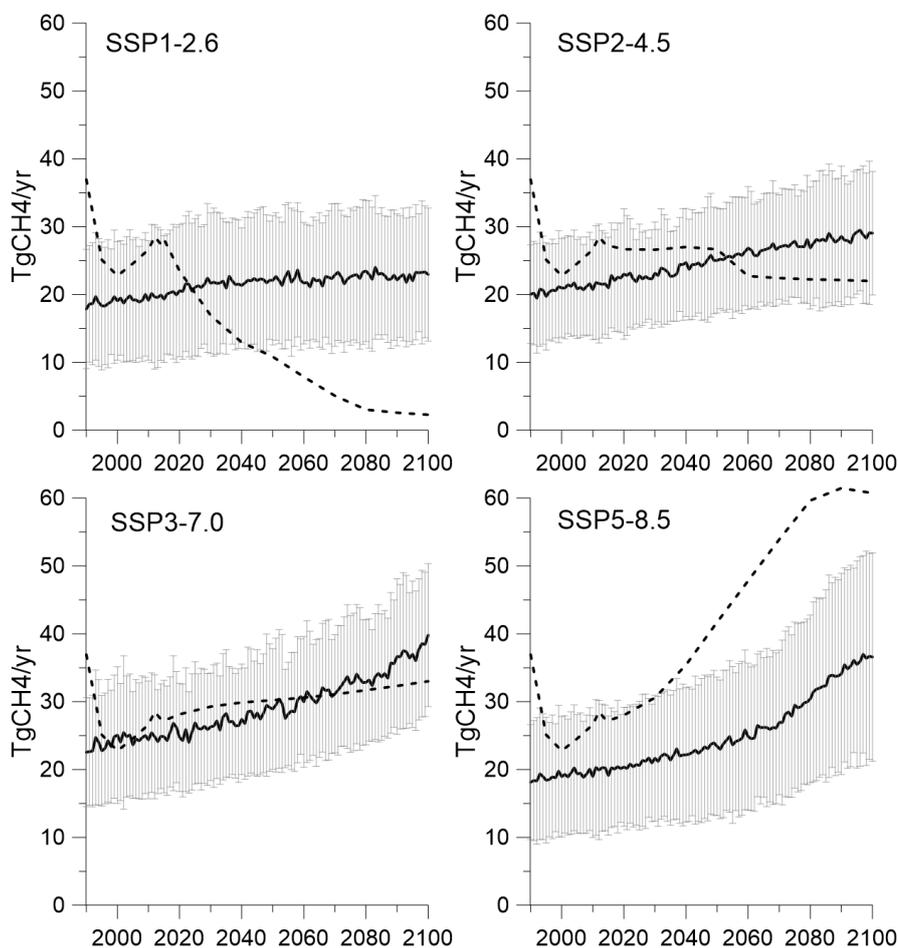

Fig. 6. $CH_4$ fluxes to the atmosphere from the territory of Russia (ensemble mean values and standard deviation) and anthropogenic emissions (dashed line) under different scenarios of anthropogenic forcing.

Estimates of natural methane emissions from the territory of Russia, as well as estimates for natural $CO_2$ fluxes, differ greatly for different CMIP6 models. Modern methane emissions (Fig. 6) range from 10 to 35 $MtCH_4$/year. Estimates using the IAP RAS ESM are closest to those of the UKESM model. Natural methane emissions into the atmosphere from the territory of Russia are comparable in magnitude with anthropogenic emissions and for all considered models under all scenarios of anthropogenic impacts increase by the end of the 21st century. According to the obtained estimates, the growth of natural methane emissions into the atmosphere in the 21st century depends on the model and scenario and may reach 300 %. A similar discrepancy in the results of model calculations for natural methane emissions into the atmosphere was obtained earlier within the framework of the WETCHIMP project (The Wetland and Wetland $CH_4$ Intercomparison of Models Project) (Melton et al., 2013). According to the results of the project, the models differ greatly in their estimates of the area of wetlands and the corresponding $CH_4$ emissions to the atmosphere, even in coordinated numerical experiments with setting the state of the atmosphere based on observational data as an external forcing.

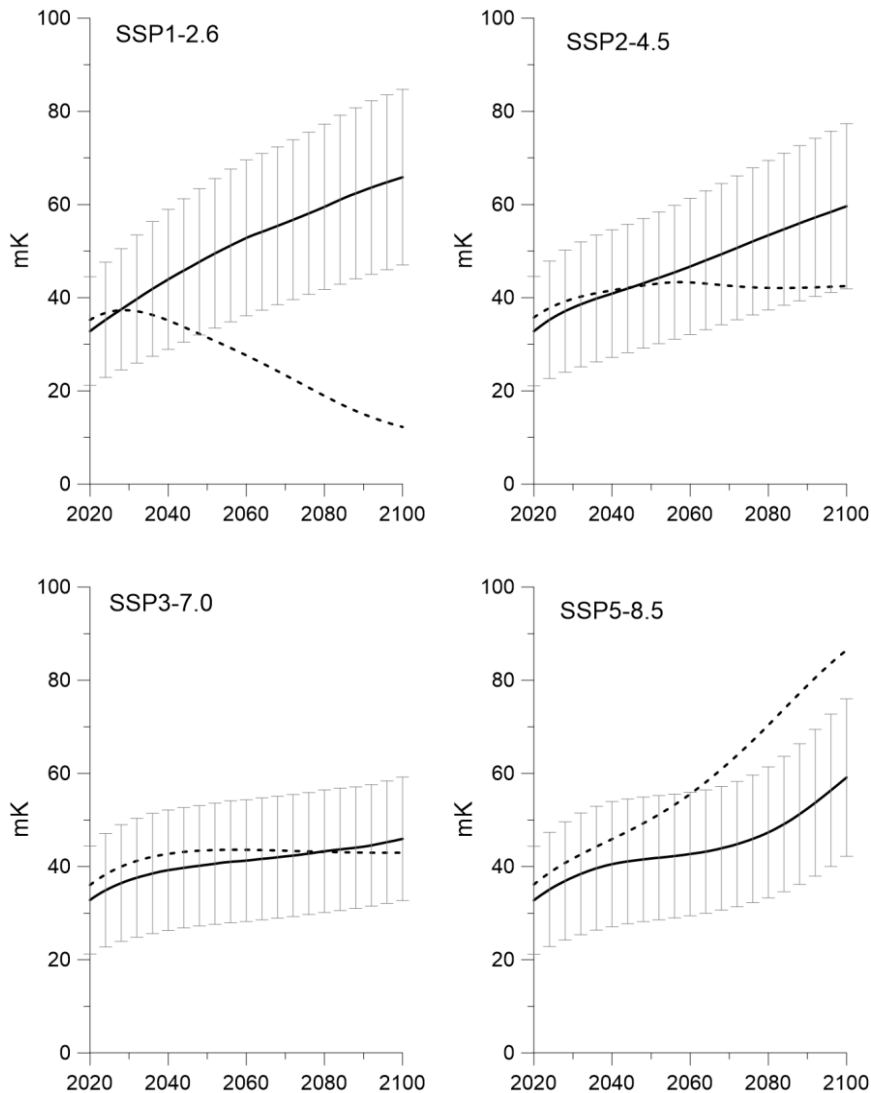

Fig. 7. Cumulative temperature potential of natural $CH_4$ fluxes [mK] to the atmosphere from the territory of Russia since 1990 as calculated with ensemble models (mean values and standard deviation) and anthropogenic fluxes (dashed line) under different scenarios of anthropogenic forcing.

The cumulative temperature potential of natural emissions of $CH_4$ into the atmosphere from the territory of Russia from 1990 to the end of the 21st century is estimated to be from 0.03 to 0.09 K, depending on the scenario of anthropogenic forcing (Fig. 7). This impact, which accelerates climate warming, is comparable in magnitude to the stabilizing impact of natural runoff from the atmosphere of $CO_2$. At the same time, according to model calculations, no tendencies to its slowing down or reduction by the end of the 21st century have been observed.

**Conclusions**

Model estimates of $CH_4$ and $CO_2$ fluxes associated with natural ecosystems of Russian regions are presented in comparison with anthropogenic greenhouse gas emissions under different scenarios of the SSP family for the 21st century. According to the results obtained, the differences in greenhouse gas fluxes as estimated using different modern models remain quite large. For Russia as a whole, the range of estimates of both $CH_4$ and $CO_2$ fluxes exceeds 100% of the average values even for the modern period. At the same time, there are general trends for changes in these fluxes. Thus, natural methane emissions to the atmosphere for all models and scenarios of anthropogenic impacts are increasing throughout the 21st century. Reaching the maximum of $CO_2$ uptake in the 21st century and its further reduction by the end of the century, revealed earlier by calculations with the IAP RAS ESM (Denisov et al., 2019), is typical for most models of the CMIP6 ensemble.

Current anthropogenic CO2 emissions from the territory of Russia are comparable in modulus to natural uptake by terrestrial ecosystems and can be largely compensated. In the SSP family scenarios, the estimates of possible future anthropogenic $CO_2$ emissions for Russia have been significantly reduced compared to the previous generation of RCP family scenarios. Therefore, compared to (Denisov et al., 2019), $CO_2$ uptake by terrestrial ecosystems of Russia in Russian regions as estimated on the basis of calculations with many models can compensate for anthropogenic emissions from the territory of Russia in the 21st century under the SSP2-4.5 scenario, and in some cases under the SSP3-7.0 scenario.

**Acknowledgements:**

This study was supported by the Russian Science Foundation (project 23-47-00104) using the results obtained under Agreement No. 075-15-2021-577 of the Ministry of Science and Higher Education of the Russian Federation with A.M. Obukhov Institute of atmospheric physics of the Russian Academy of Sciences.